\documentclass[prl,superscriptaddress,twocolumn]{revtex4}

\usepackage{hyperref}
\usepackage{ifthen}
\usepackage{graphics}

\newcommand{\psec}[1]{\emph{#1.}---}

\newcommand{\hmpc}{h/{\rm Mpc}}

\begin{document}

\title{Unscreening Modified Gravity in the Matter Power Spectrum}

\author{Lucas~Lombriser}
\affiliation{Institute for Astronomy, University of Edinburgh, Royal Observatory, Blackford Hill, Edinburgh EH9~3HJ, U.K.}
\author{Fergus~Simpson}
\affiliation{ICC, University of Barcelona (UB-IEEC), Marti~i~Franques~1, 08028, Barcelona, Spain}
\author{Alexander~Mead}
\affiliation{Institute for Astronomy, University of Edinburgh, Royal Observatory, Blackford Hill, Edinburgh EH9~3HJ, U.K.}

\date{\today}

\begin{abstract}
Viable modifications of gravity that may produce cosmic acceleration need to be screened in high-density regions such as the Solar System, where general relativity is well tested.
Screening mechanisms also prevent strong anomalies in the large-scale structure and limit the constraints that can be inferred on these gravity models from cosmology.
We find that by suppressing the contribution of the screened high-density regions in the matter power spectrum, allowing a greater contribution of unscreened low densities, modified gravity models can be more readily discriminated from the concordance cosmology.
Moreover, by variation of density thresholds, degeneracies with other effects may be dealt with more adequately.
Specializing to chameleon gravity as a worked example for screening in modified gravity, employing $N$-body simulations of $f(R)$ models and the halo model of chameleon theories, we demonstrate the effectiveness of this method.
We find that a percent-level measurement of the clipped power at $k<0.3~\hmpc$ can yield constraints on chameleon models that are more stringent than what is inferred from Solar System tests or distance indicators in unscreened dwarf galaxies.
Finally, we verify that our method is also applicable to the Vainshtein mechanism.

\end{abstract}

\maketitle

\psec{Introduction}
%
Determining the nature of the accelerated expansion of our Universe is a prime endeavor to cosmologists.
In the conventional picture, the flat $\Lambda$ cold dark matter ($\Lambda$CDM) concordance model based on general relativity (GR), a cosmological constant $\Lambda$ contributes the bulk of the present energy density in the cosmos and drives the late-time acceleration.
While alternatively, a modification of gravity may be responsible for cosmic acceleration, stringent limitations from experiments within our Solar System must be satisfied.
A number of screening mechanisms~\cite{vainshtein:72,khoury:03a,babichev:09,hinterbichler:10,lombriser:14b} have been identified that can suppress modifications of gravity in high-density regions to recover GR, while still generating significant modifications within lower densities on larger, cosmological scales.
However, this suppression effect, along with other nonlinear effects, also prevents strong anomalies from manifesting in the averaged large-scale structure of our Universe~\cite{oyaizu:08} and limits the constraints that can be inferred on these gravity models from cosmology.

Given the density dependence of the screening effect, in this Letter, we propose the downweighting of high-density regions in statistical observables such as the matter power spectrum $P(k)$ to enhance, or unscreen,
the signatures of modified gravity and improve observational constraints.
Such a weighting is conducted in the clipping method of Ref.~\cite{simpson:11},
with the original motivation of facilitating the modeling of $P(k)$ by reducing contributions of high densities, where the assumptions of perturbation theory break down.
As a worked example, we first focus on Hu-Sawicki~\cite{hu:07a} $f(R)$ gravity~\cite{buchdahl:70}, which employs the chameleon screening mechanism~\cite{khoury:03a}.
We analyze effects on the power spectrum from clipping density fields in numerical simulations of the model.
Using the halo model of chameleon theories~\cite{lombriser:13c}, we then generalize our findings to chameleon models with arbitrary gravitational coupling and exponents of the chameleon field potential. 
We also verify the applicability of~clipping~to~Vainshtein~screening~\cite{vainshtein:72}.

\psec{Chameleon gravity}
%
We first specialize to the Hu-Sawicki $f(R)$ ($n=1$) model, where the nonlinear function $f(R) \simeq -2\Lambda - f_{R0}\bar{R}_0^2/R$ of the Ricci scalar $R$ is added to the Einstein-Hilbert action.
Here, bars denote quantities evaluated at the cosmological background, zeros refer to present time, $f_R \equiv df/dR$ is the additional, scalar degree of freedom of the model, and $f_{R0}\equiv\bar{f}_{R}(z=0)$.
In the quasistatic approximation and for $|f_{R0}|\ll1$, the modified Poisson equation becomes (see, e.g., Refs.~\cite{capozziello:11a,capozziello:11b})
\begin{equation}
 \nabla^2\Psi = \frac{16\pi\,G}{3}\delta \rho_{\rm m} - \frac{1}{6}\delta R(f_R), \label{eq:fRpoisson}
\end{equation}
where $\delta$ denotes perturbations with respect to the cosmological background, e.g., $\delta R = R-\bar{R}$, and $\Psi\equiv\delta g_{00}/(2g_{00})$.
The scalar field equation is given by
\begin{equation}
 \nabla^2\delta f_R = - \frac{8\pi G}{3} \delta \rho_{\rm m} + \frac{1}{3}  \delta R\left(f_R\right). \label{eq:fRfield}
\end{equation}
The chameleon mechanism works such that in a high-density region $\delta R\simeq8\pi G\delta \rho_{\rm m}$ and hence Eq.~(\ref{eq:fRpoisson}) reduces to the Poisson equation of Newtonian gravity.
In contrast, at low densities $\delta R \simeq \left.\partial R/\partial f_R\right|_{R=\bar{R}} = 3m^2\delta f_R$, which when applied to Eqs.~(\ref{eq:fRpoisson}) and (\ref{eq:fRfield}) in Fourier space yields the unscreened modified Poisson equation
\begin{equation}
 k^2 \hat{\Psi} = - 4\pi\,G \left\{ \frac{4}{3} - \frac{1}{3} \left[ \left( \frac{k}{m \, a} \right)^2 + 1 \right]^{-1} \right\} a^2 \widehat{\delta \rho}_{\rm m}. \label{eq:fRpoissonlin}
\end{equation}
Hence, whereas in high-density regions gravity returns to Newtonian due to the chameleon mechanism, at low densities and scales below $m^{-1}$, gravitational interactions remain enhanced by a factor of $4/3$.
In particular, Eq.~({\ref{eq:fRpoissonlin}}) applies to the linear perturbation regime.
Note that $f(R)$ models correspond to a Brans-Dicke scalar-tensor theory with Brans-Dicke parameter $\omega=0$, Jordan-frame scalar field $\varphi=1+f_R$, and scalar field potential $U=(R f_R-f)/2$.
More generally, the chameleon mechanism is realized for scalar-tensor models with scalar field potential $U(\varphi)-\Lambda \propto (1-\varphi)^{\alpha}$ with $\alpha\equiv n/(n+1) \in(0,1)$, where the gravitational coupling is maximally enhanced by a factor of $(4+2\omega)/(3+2\omega)$ with $\omega>-3/2$~\cite{lombriser:13c}.
Importantly, although serving as very useful example for screening mechanisms, chameleon models do not yield a genuine self-acceleration of the cosmic expansion due to their gravitational modifications~\cite{wang:12}.

In order to obtain accurate results in the nonlinear regime of the Hu-Sawicki $f(R)$ model, we use dark matter $N$-body simulations run in Ref.~\cite{mead:14} with the \texttt{ECOSMOG} code of Ref.~\cite{li:11}, which uses particles and adaptive meshes to solve Eqs.~(\ref{eq:fRpoisson}) and (\ref{eq:fRfield}).
The background expansion is taken to be equivalent to that of $\Lambda$CDM, appropriate for observationally interesting $f_{R0}$ values.
We use simulations of the concordance model and $f(R)$ gravity where $|f_{R0}|$ is $10^{-4}$ (F4), $10^{-5}$ (F5), and $10^{-6}$ (F6), all sharing an initial seed and cosmological parameters.
Each simulation contains $512^3$ particles in a box with $L=512h^{-1} \mathrm{Mpc}$; $h = 0.697$, $\Omega_\mathrm{m} =0.281$, $\Omega_\mathrm{b} = 0.046$, $\Omega_{\Lambda} = 0.719$, $n_\mathrm{s} = 0.971$, and amplitude of the matter power spectrum such that $\sigma_8 = 0.82$ in $\Lambda$CDM.
Note that $\sigma_8$ is larger for $f(R)$ gravity due to the enhanced forces and growth of structure.
An initial power spectrum for the simulations was generated using \texttt{MPGRAFIC}~\cite{prunet:13}.
The particle mass in each case is $\simeq {7.80}\times 10^{10}h^{-1}\,\mathrm{M_\odot}$.
Each simulation has exactly the same initial power spectrum $(z_\mathrm{i}=49)$ and differences between models are confined to different strengths of enhanced perturbation growth at late times and different strengths of screening.
For the extrapolation of nonlinear physics from $f(R)$ gravity to more general chameleon models with $\omega\neq0$, we employ the halo model of chameleon theories developed in Ref.~\cite{lombriser:13c}, which accounts for the density dependence of the effective gravitational coupling in the nonlinear regime and provides matter power spectra that are in good agreement with measurements in $f(R)$ $N$-body simulations (see Fig.~4 in Ref.~\cite{lombriser:13c}).
%

\psec{Clipping the density fields}
%
The spatial distribution of matter on cosmological scales may be quantified by the fractional overdensity field $\delta(x) \equiv \rho_{\rm m}/\bar{\rho}_{\rm m} - 1$.
We construct the density fields from the simulations using a cloud-in-cell interpolation on a $256^3$ Cartesian mesh in each cell.
Clipping is a local density transformation characterized by enforcing a maximum fractional overdensity $\delta_0$ such that~\cite{simpson:11}
\begin{equation}
 \delta_c(x) =
 \left\{ \begin{array}{ll} 
  \delta_0, & \delta(x)>\delta_0, \\
  \delta(x), & \delta(x)\leq\delta_0.
 \end{array} \right. 
 \label{eq:clip}
\end{equation}
We shall also make use of applying a minimum instead of a maximum threshold, which is equivalent to clipping the negative field $-\delta (x)$.
While this may prove more challenging to apply to real data, due to the lower signal to noise associated with cosmic voids, it will serve as a useful validation test in our simulations and support for the concept of weighting to unscreen or screen gravitational modifications.
We quantify the clipping strength in terms of the fractional loss of power in the lowest $k$-bin, applying a simple iterative procedure to determine the threshold $\delta_0$ required to establish the desired fraction.
Defining clipping strength directly in terms of $\delta_0$ is another possibility,
but complicates the comparison of results from fields with a different choice of smoothing length.

\begin{figure*}
 \centering
 \resizebox{\hsize}{!}{
  \resizebox{\hsize}{!}{\includegraphics{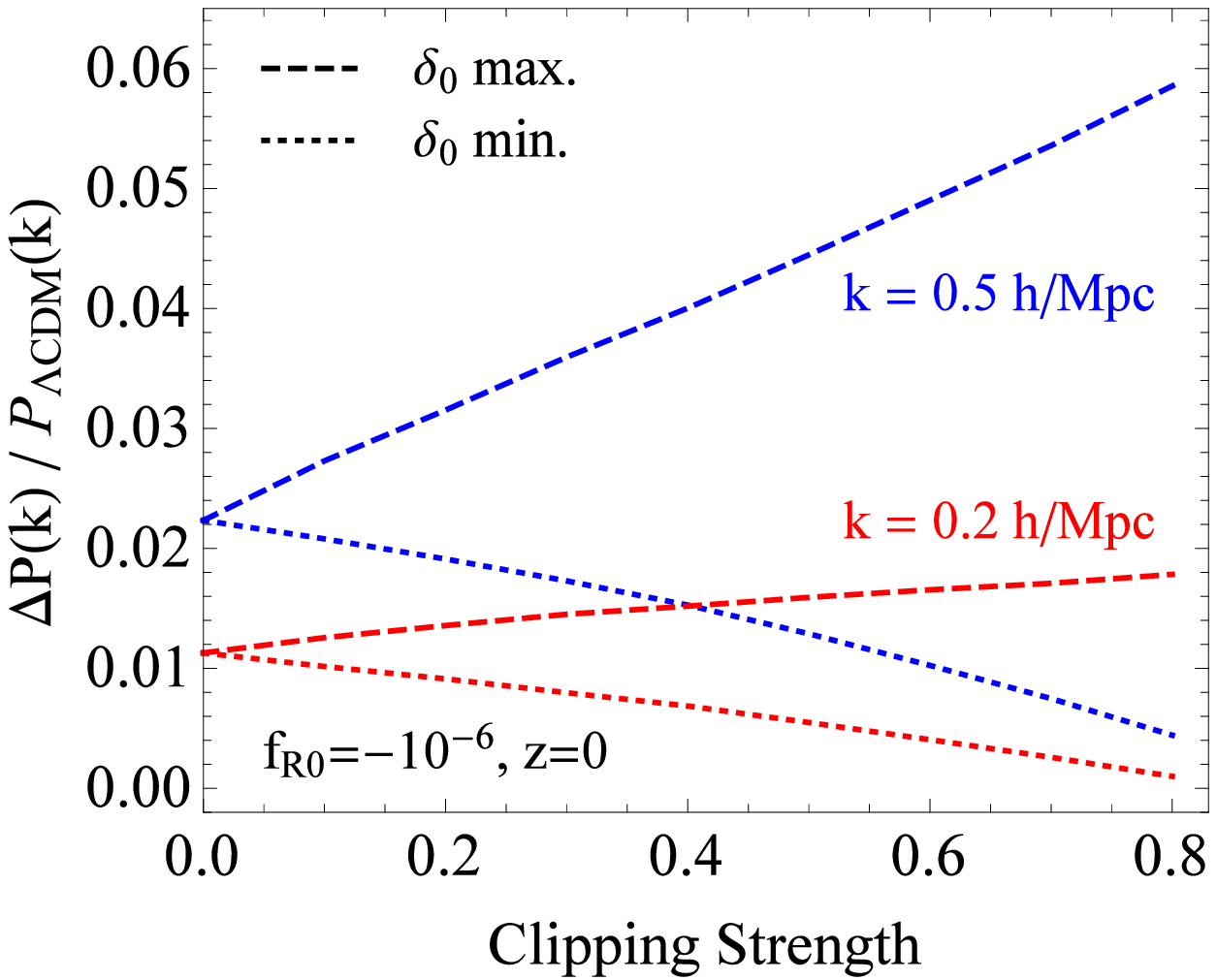}}
  \resizebox{0.971\hsize}{!}{\includegraphics{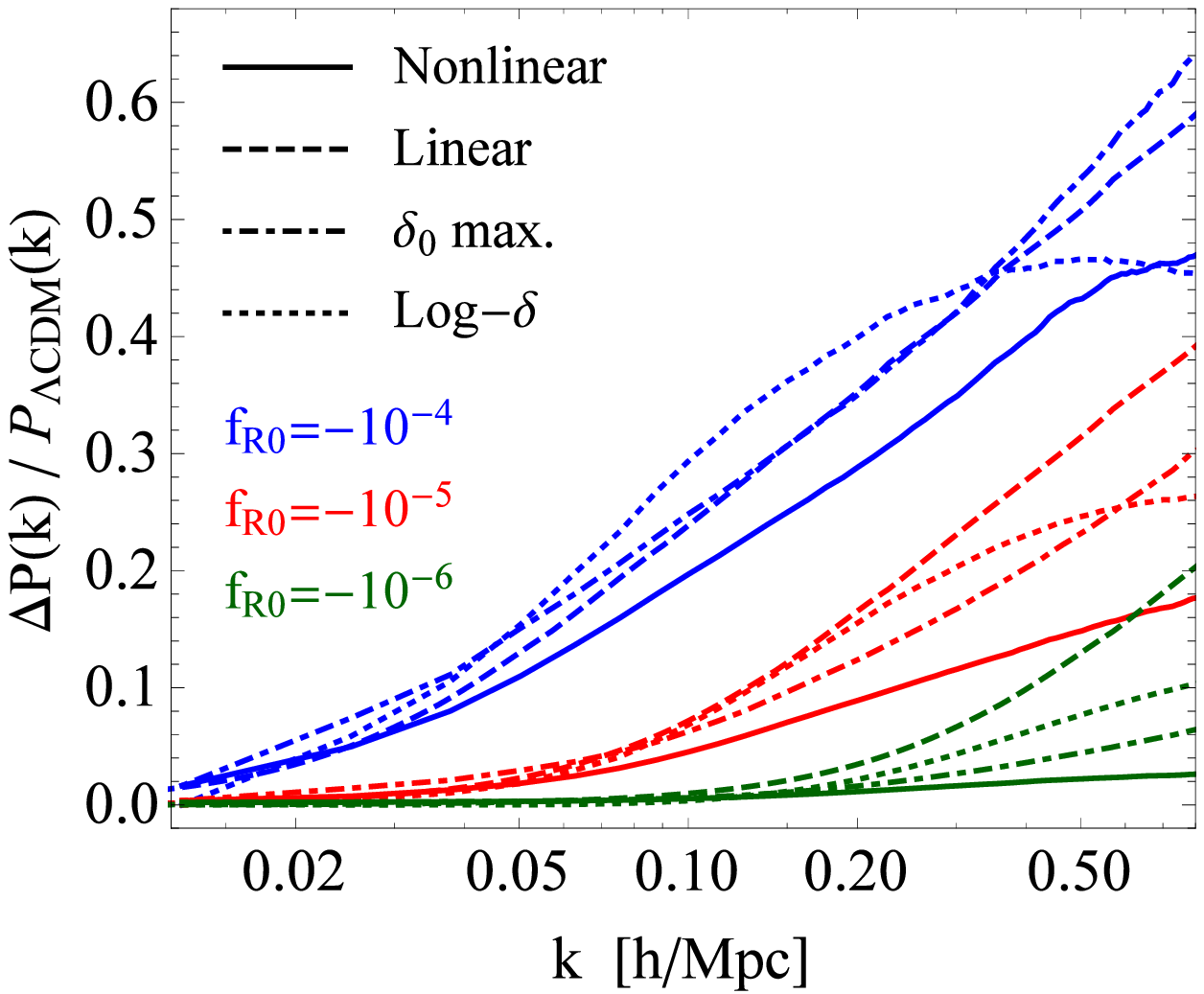}}
  \resizebox{0.976\hsize}{!}{\includegraphics{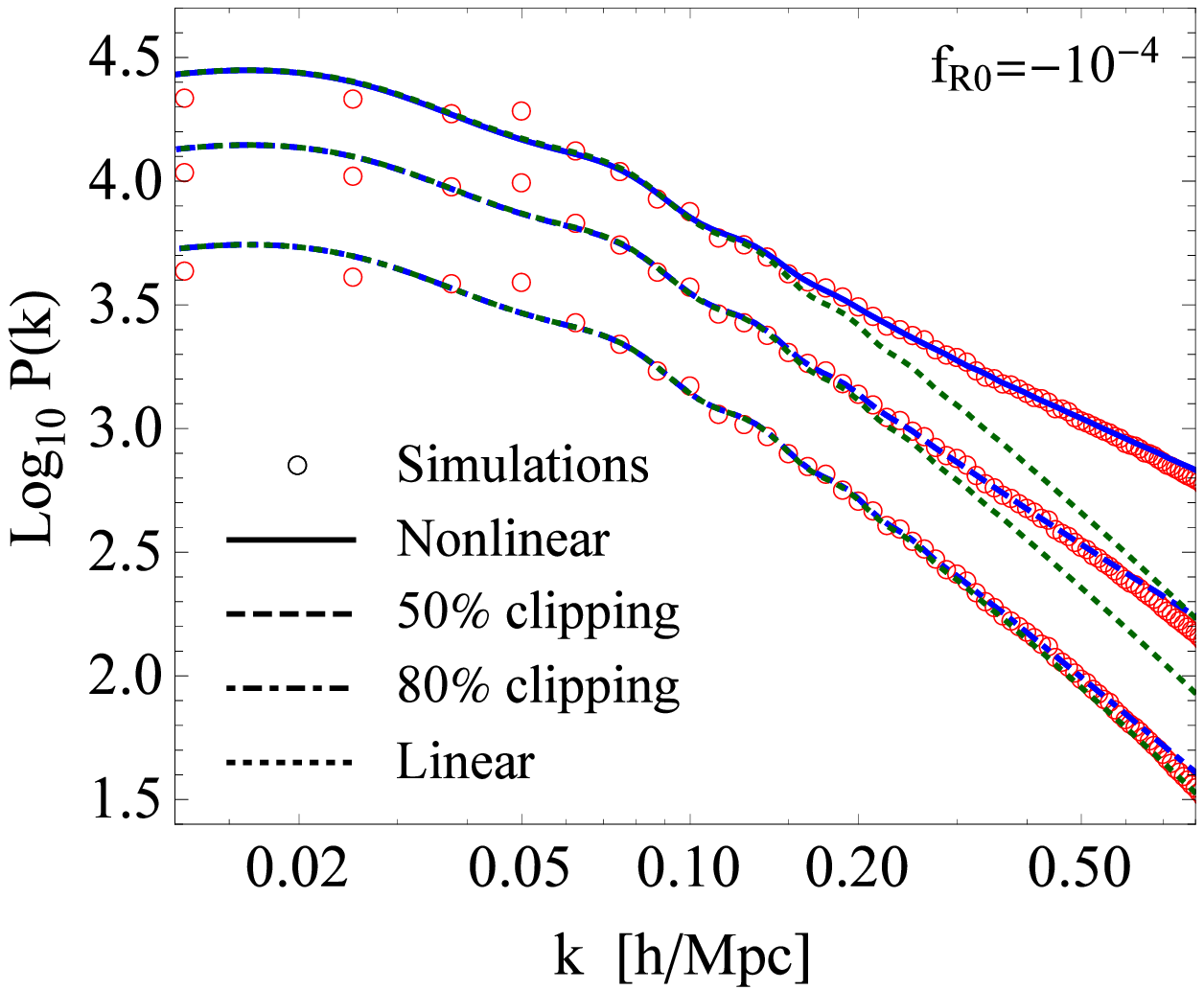}}
 }
\caption{\textit{Left:} Fractional difference between simulated matter power spectra of $f(R)$ gravity $(|f_{R0}|=10^{-6})$ and $\Lambda$CDM at two different $k$-bins.
By clipping high densities, contributions from self-screened regions are downweighted and unscreened features enhanced (dashed curves).
In the reverse case of applying a minimum threshold to the density field, contributions from unscreened low densities are downweighted, enhancing screened regions, and the power spectra converge (dotted lines).
\textit{Middle:} Fractional departure from the $\Lambda$CDM power spectrum in three different $f(R)$ models for unclipped fields (solid lines), linear perturbation theory (dashed lines), clipped fields with power at the largest scales reduced by $50\%$ (dot-dashed curves), and logarithmic density transformation with matching to the $\Lambda$CDM power at the lowest $k$-bin (dotted curves).
\textit{Right:} Clipped $f(R)$ power spectra from simulations (circles) compared to weighting linear and one-loop contributions (dashed curves).
Increased clipping strength enhances the weight of the unscreened linear (dotted curves) relative to the one-loop power.
\label{fig:one}}
\end{figure*}

Given the nature of the chameleon mechanism, the clipping transformation promises the extraction of more information on the gravitational physics, as it allows us to focus on the less dense, unscreened regions of the Universe, where there exists a greater difference from GR. 
In the left panel of Fig.~\ref{fig:one}, we compare the deviation in power between the F6 and $\Lambda$CDM simulations as a function of clipping strength.
With increased clipping strength, the difference in $P(k)$ is enhanced.
At $k=0.5~\hmpc$ this difference more than doubles, greatly facilitating the discrimination between the two models.
In practice, the amount of clipping that can realistically be applied will depend on the noise levels, limited by the number density of galaxies. 
To see the reverse of this effect, we also apply the clipping transformation to the negative field $-\delta(x)$.
Since this is clipping voids, this reduces the contributions from unscreened regions to $P(k)$.
Therefore, as expected, stronger clipping leads to convergence between the modified gravity and $\Lambda$CDM density fields.  
Note that we do not expect a full unscreening of modified dynamics by clipping high-density peaks since this only affects self-screening but not environmental screening~\cite{lombriser:13c}.
In the opposite case of clipping low-density troughs, however, only self-screened regions with GR dynamics remain.

In the middle panel of Fig.~\ref{fig:one}, we show the ratios of the power spectra of F4-6 to that of $\Lambda$CDM simulations and the corresponding ratios after clipping has been applied such that the large-scale power is reduced by $50\%$.
We also show the results of instead applying a logarithmic density transform~\cite{neyrinck:09}.
Both local transformations suppress contributions from the densest regions of the field.
Note that while the shape recovered from the  logarithmic transform is closer to linear theory, for sparsely sampled fields the logarithmic transform becomes unstable whereas clipping is insensitive to the number density of sources~\cite{simpson:13}.
Regarding the proximity to linear theory, Ref.~\cite{simpson:13} demonstrated that the clipped $\Lambda$CDM matter power spectrum is well described by a linear combination of the linear and one-loop contributions, where higher-order terms are strongly suppressed.
Thereby, increased clipping downweights the one-loop relative to the linear contribution.
We verify the recovery of linear theory for $f(R)$ gravity in the right-hand panel of Fig.~\ref{fig:one}, using F4 simulations, where chameleon screening affects $P(k)$ at $k\gtrsim0.1~\hmpc$ (see Fig.~2 in Ref.~\cite{oyaizu:08}).
With increased clipping, $P(k)$ more closely reflects linear, unscreened, theory with a reduced relative weight of the one-loop correction.

\psec{Differential clipping to break degeneracies}
%
Effects on the total matter power spectrum from $f(R)$ modifications of gravity, and hence other chameleon models, have been shown to yield some degeneracy with effects from varying the total neutrino mass~\cite{motohashi:10, baldi:13} or from baryonic feedback processes~\cite{puchwein:13}.
It is important to note that while degenerate in $P(k)$, these contributions are not affecting a particular range of densities in the same manner as the chameleon modification.
More precisely, nonvanishing neutrino masses suppress $P(k)$ predominantly linearly, i.e., with negligible preference on density, and baryonic feedback processes emanate from high-density regions.
Hence, a combined analysis of clipped power spectra employing different density thresholds can break degeneracies between the different contributions.
Furthermore, in a parameter estimation analysis of chameleon models, we need to allow for a variation of cosmological parameters.
In particular the growth of structure can be strongly altered by variations of $\sigma_8$ or $\Omega_{\rm m}$.
Using \texttt{HALOFIT}~\cite{smith:02}, we estimate effects of varying these parameters on the clipped power by comparing signatures in the linear and nonlinear $P(k)$. 
While enhancing the amplitude of $P(k)$ with increasing $\sigma_8$, the shape of the nonlinear enhancement resembles that of a chameleon model.
Hence, with absent information on the absolute amplitude of $P(k)$ due to galaxy bias, there clearly is a degeneracy between variations in $\sigma_8$ and $f_{R0}$.
However, the $\Lambda$CDM case with a larger value of $\sigma_8$, will experience a strong reduction of the enhancement in power at large $k$ after clipping.
In contrast, the $f(R)$ power spectrum increases the enhancement at large $k$ after clipping, becoming more linear and unscreened. 
Therefore variations in $f_{R0}$ and $\sigma_8$ respond to clipping in a qualitatively different manner, breaking the degeneracy.
Modifications in the shape of $P(k)$ due to changes in $\Omega_{\rm m}$ are qualitatively different from $f(R)$ modifications, e.g., changing baryon acoustic oscillation features.
Although a partial suppression in the change of power attributed to $\Omega_{\rm m}$ variations is seen in nonlinear compared to linear theory, it does not reproduce the strong screening effect of chameleon models.
Finally, note that since focusing on differences in the shape of $P(k)$, effects of linear galaxy bias can be neglected.
However, through redshift-space distortions, in chameleon models the ratio between galaxy and dark matter density becomes scale dependent.
Since adding to the deviations between the shape of modified and $\Lambda$CDM galaxy power spectra~\cite{lombriser:13a}, we conservatively assume this ratio to be constant when estimating potential observational bounds on chameleon models.

\psec{Outperforming Solar System constraints}
%
The requirement that the Milky Way dark matter halo screens the Solar System sets a constraint on the chameleon field amplitude of $|\bar{\varphi}_0-1|\lesssim 5\times10^{-6}/(6+4\omega)$~\cite{lombriser:13c}.
Similarly strong constraints can be obtained from the absence of deviations in luminosity distances from different types of distance indicators in unscreened dwarf galaxies~\cite{jain:12}.
Current cosmological constraints are about 2 orders of magnitude weaker than Solar System bounds~\cite{terukina:13,dossett:14,planck:15}. 
We refer to Ref.~\cite{lombriser:14a} for a review of constraints on chameleon gravity.
Having analyzed the effect of clipping on the matter power spectrum, we estimate the constraints that can be inferred from applying this method to observations of galaxy clustering.
While fractional errors in the measurement can be kept approximately constant, clipping reduces systematic errors from modeling uncertainties of the nonlinear structure contributing at large $k$ such that constraints on cosmological parameters can be improved (see Ref.~\cite{simpson:15}).
We assume that a future measurement of the clipped $P(k)$ can discriminate 1\% deviations from the fiducial $\Lambda$CDM shape at $k\leq0.3~\hmpc$.
Note that this is a conservative estimate with surveys already yielding subpercent-level measurements of the acoustic features~\cite{anderson:13}.
We compute the modified nonlinear power employing the halo model of chameleon theories~\cite{lombriser:13c}. 
As clipping chameleon densities mainly removes regions where GR is recovered, we approximate the clipped chameleon power spectrum by removing the difference of unclipped to clipped power obtained from the $\Lambda$CDM simulation.
For $k\leq0.3~\hmpc$, this simple approach reproduces the absolute clipped power for F4 and F6 at the few percent and permille level, respectively, and recovers the measured fractional difference to $\Lambda$CDM within 20\%, increasingly underestimating it from F4 to F6.
Employing this method and varying the background field value $\bar{\varphi}_0$, the coupling strength set by $\omega$, and the exponent of the scalar field potential $\alpha$, we set constraints where the deviation between the clipped chameleon and $\Lambda$CDM $P(k)$ at $k=0.3~\hmpc$ exceeds $1\%$.

\begin{figure}
 \centering
  \resizebox{0.7\hsize}{!}{\includegraphics{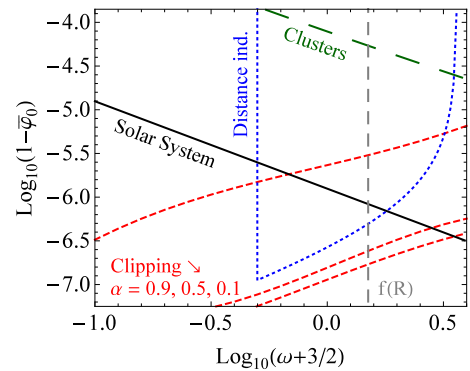}}
\caption{Forecast of upper bounds on the chameleon field amplitude $1-\bar{\varphi}_0$ from the clipped matter power spectrum as a function of gravitational coupling strength set by the Brans-Dicke parameter $\omega$ and for different values of the exponent of the chameleon field potential $\alpha$. The constraints are compared to upper bounds inferred from Solar System~\cite{lombriser:14c}, astrophysical distance indicator~\cite{jain:12}, and cosmological (cluster) observations~\cite{terukina:13} (cf.~Ref.~\cite{lombriser:14a}).
For a range of chameleon model parameters, clipping can provide the strongest constraints.
\label{fig:constraints}}
\end{figure}

We present our results in Fig.~\ref{fig:constraints}, comparing them to Solar System, astrophysical distance indicator, and current cosmological bounds.
We find that for the simulated Hu-Sawicki model ($\alpha=0.5$), clipping constraints on $f_{R0}$ can improve upon existing cosmological constraints, using clusters~\cite{terukina:13}, the matter power spectrum~\cite{dossett:14}, or cosmic microwave background with redshift-space distortion data~\cite{planck:15}, by 2--3 orders of magnitude and outperform the Solar System and astrophysical bounds as well as local tests of the equivalence principle~\cite{khoury:03a,lombriser:14a,capozziello:07}.
For smaller values of $\omega$, corresponding to larger force modifications, the improvement of constraints from clipping over Solar System bounds is even greater.
This is not surprising since larger force modifications also imply a more efficient screening of the Solar System region whereas gravitational dynamics in unscreened low densities is modified even more strongly.
Constraints in this region of parameter space may also confirm or rule out chameleon models as an explanation of the observed cored density profiles of dwarf spheroidal galaxies in the Milky Way~\cite{lombriser:14c}.
In the opposite limit of increasing $\omega$, i.e., weakening gravitational coupling, Solar System bounds strengthen and surpass the decreasing clipping constraints.
Importantly, these power spectrum constraints clearly depend on the exponent of the scalar field potential $\alpha$.

\psec{Vainshtein screening}
%
A similar density dependence to the chameleon mechanism enters the effective gravitational coupling at nonlinear scales in models exhibiting a Vainshtein~\cite{vainshtein:72} screening effect~\cite{schmidt:09}.
While the modifications of $P(k)$ are comparable, the Vainshtein mechanism originates from derivative self-interactions in contrast to the scalar field potential in chameleon gravity, causing some differences in the screening behavior~\cite{falck:15}. Chameleon screening depends on the gap between an object's internal and exterior scalar field values set by the corresponding densities. Hence, an object can be self-screened or environmentally screened by a large local or environmental density, respectively. In contrast, in the Vainshtein mechanism, screening depends on morphology of the source but not on its environment, introducing a large screening scale around an object, suppressing modifications to its own gravitational field but still responding to external field modifications.
We verify that the clipping method is also applicable to the Vainshtein mechanism by employing 50\% clipping to the nDGP~\cite{dvali:00} simulations of Ref.~\cite{falck:15}.
We find an enhancement of $\Delta P/P$ (cf. Fig.~\ref{fig:one}) of 70\%, 60\%, and 20\% at screened scales of $k\sim1.5~\hmpc$ for models with $\sigma_8$ values equal to the F4, F5, and F6 $f(R)$ scenarios, respectively.
We also expect our method to be applicable to further screening mechanisms like the symmetron~\cite{hinterbichler:10} or k-mouflage~\cite{babichev:09}, however, not to linear shielding~\cite{lombriser:14b}.

\psec{Conclusions}
%
Modifications of gravity potentially explaining cosmic acceleration need to employ a screening mechanism that allows modifications in low densities at large scales while suppressing them in high-density regions such as the Solar System.
Such screening mechanisms also suppress anomalies in the averaged cosmological large-scale structure, limiting observational constraints
inferred on such models.
By clipping the density fields at a maximal threshold, contributions of screened high-density regions to the matter power spectrum can be downweighted, enhancing modified gravity effects.
We applied this method to chameleon models
with particular emphasis on $f(R)$ gravity to demonstrate that it can improve cosmological constraints on the models to a level stronger than the currently most stringent bounds from Solar System and astrophysical tests.
We also verified that clipping is applicable to Vainshtein screening and expect it to be employable for further nonlinear screening mechanisms.

We thank Baojiu Li, Kazuya Koyama, and Gong-Bo Zhao for sharing $f(R)$ and nDGP $N$-body simulations with us.
L.L.~is supported by the U.K.~STFC Consolidated Grant for Astronomy and Astrophysics at the University of Edinburgh.
A.M.~ and F.S.~acknowledge support from the European Research Council under the EC FP7 Grants No.~240185 and No.~240117, respectively. 
Please contact the authors for access to research materials.


\bibliographystyle{arxiv_physrev_mod}
\bibliography{MGclip}

\def\eprinttmppp@#1arXiv:@{#1}
\providecommand{\arxivlink[1]}{\href{http://arxiv.org/abs/#1}{arXiv:#1}}
\def\eprinttmp@#1arXiv:#2 [#3]#4@{\ifthenelse{\equal{#3}{x}}{\ifthenelse{
\equal{#1}{}}{\arxivlink{\eprinttmppp@#2@}}{\arxivlink{#1}}}{\arxivlink{#2}
  [#3]}}
\providecommand{\eprintlink}[1]{\eprinttmp@#1arXiv: [x]@}
\renewcommand{\eprint}[1]{\eprintlink{#1}}
\providecommand{\eprintmod}[1][XXXX.XXXX]{\eprintlink{#1}}
\providecommand{\adsurl}[1]{\href{#1}{ADS}}
\renewcommand{\bibinfo}[2]{\ifthenelse{\equal{#1}{isbn}}{\href{http://cosmolog%
ist.info/ISBN/#2}{#2}}{#2}}
\begin{thebibliography}{36}
\expandafter\ifx\csname natexlab\endcsname\relax\def\natexlab#1{#1}\fi
\expandafter\ifx\csname bibnamefont\endcsname\relax
  \def\bibnamefont#1{#1}\fi
\expandafter\ifx\csname bibfnamefont\endcsname\relax
  \def\bibfnamefont#1{#1}\fi
\expandafter\ifx\csname citenamefont\endcsname\relax
  \def\citenamefont#1{#1}\fi
\expandafter\ifx\csname url\endcsname\relax
  \def\url#1{\texttt{#1}}\fi
\expandafter\ifx\csname urlprefix\endcsname\relax\def\urlprefix{URL }\fi

\bibitem{vainshtein:72}
A.~Vainshtein,
\newblock Phys. Lett. {\bf B39}, 393 (1972).

\bibitem{khoury:03a}
J.~Khoury and A.~Weltman,
\newblock Phys. Rev. Lett. {\bf 93}, 171104 (2004).

\bibitem{babichev:09}
E.~{Babichev}, C.~{Deffayet} and R.~{Ziour},
\newblock IJMPD {\bf 18}, 2147 (2009).

\bibitem{hinterbichler:10}
K.~Hinterbichler and J.~Khoury,
\newblock Phys. Rev. Lett. {\bf 104}, 231301 (2010).

\bibitem{lombriser:14b}
L.~Lombriser and A.~Taylor,
\newblock Phys.Rev.Lett. {\bf 114}, 031101 (2015).

\bibitem{oyaizu:08}
H.~Oyaizu, M.~Lima and W.~Hu,
\newblock Phys. Rev. {\bf D78}, 123524 (2008).

\bibitem{simpson:11}
F.~{Simpson}, J.~B. {James}, A.~F. {Heavens} and C.~{Heymans},
\newblock Phys. Rev. Lett. {\bf 107}, A261301 (2011).

\bibitem{hu:07a}
W.~Hu and I.~Sawicki,
\newblock Phys. Rev. {\bf D76}, 064004 (2007).

\bibitem{buchdahl:70}
H.~A. {Buchdahl},
\newblock Mon. Not. Roy. Astron. Soc. {\bf 150}, 1 (1970).

\bibitem{lombriser:13c}
L.~Lombriser, K.~Koyama and B.~Li,
\newblock JCAP {\bf 1403}, 021 (2014).

\bibitem{capozziello:11a}
S.~Capozziello, M.~De~Laurentis, S.~Odintsov and A.~Stabile,
\newblock Phys.Rev. {\bf D83}, 064004 (2011).

\bibitem{capozziello:11b}
S.~Capozziello, M.~De~Laurentis, I.~De~Martino, M.~Formisano and S.~Odintsov,
\newblock Phys.Rev. {\bf D85}, 044022 (2012).

\bibitem{wang:12}
J.~Wang, L.~Hui and J.~Khoury,
\newblock Phys. Rev. Lett. {\bf 109}, 241301 (2012).

\bibitem{mead:14}
A.~Mead, J.~Peacock, L.~Lombriser and B.~Li,
\newblock Mon. Not. Roy. Astron. Soc. {\bf 452}, 4203 (2015).

\bibitem{li:11}
B.~{Li}, G.-B. {Zhao}, R.~{Teyssier} and K.~{Koyama},
\newblock JCAP {\bf 1}, 51 (2012).

\bibitem{prunet:13}
S.~{Prunet} and C.~{Pichon},
\newblock {MPgrafic: A parallel MPI version of Grafic-1}, 2013,
\newblock Astrophysics Source Code Library, ascl:1304.014.

\bibitem{neyrinck:09}
M.~C. Neyrinck, I.~Szapudi and A.~S. Szalay,
\newblock Astrophys. J. {\bf 698}, L90 (2009).

\bibitem{simpson:13}
F.~{Simpson}, A.~F. {Heavens} and C.~{Heymans},
\newblock Phys. Rev. {\bf D88}, 083510 (2013).

\bibitem{motohashi:10}
H.~Motohashi, A.~A. Starobinsky and J.~Yokoyama,
\newblock Prog. Theor. Phys. {\bf 124}, 541 (2010).

\bibitem{baldi:13}
M.~Baldi {\em et~al.},
\newblock Mon. Not. Roy. Astron. Soc. {\bf 440}, 75 (2014).

\bibitem{puchwein:13}
E.~{Puchwein}, M.~{Baldi} and V.~{Springel},
\newblock Mon. Not. Roy. Astron. Soc. {\bf 436}, 348 (2013).

\bibitem{smith:02}
Virgo Consortium, R.~Smith {\em et~al.},
\newblock Mon. Not. Roy. Astron. Soc. {\bf 341}, 1311 (2003).

\bibitem{lombriser:13a}
L.~Lombriser, J.~Yoo and K.~Koyama,
\newblock Phys.Rev. {\bf D87}, 104019 (2013).

\bibitem{jain:12}
B.~Jain, V.~Vikram and J.~Sakstein,
\newblock Astrophys. J. {\bf 779}, 39 (2013).

\bibitem{terukina:13}
A.~Terukina {\em et~al.},
\newblock JCAP {\bf 1404}, 013 (2014).

\bibitem{dossett:14}
J.~Dossett, B.~Hu and D.~Parkinson,
\newblock JCAP {\bf 1403}, 046 (2014).

\bibitem{planck:15}
Planck, P.~Ade {\em et~al.},
\newblock \eprintmod[arXiv:1502.01590].

\bibitem{lombriser:14a}
L.~Lombriser,
\newblock Annalen Phys. {\bf 526}, 259 (2014).

\bibitem{simpson:15}
F.~Simpson {\em et~al.},
\newblock \eprintmod[arXiv:1505.03865].

\bibitem{anderson:13}
L.~{Anderson} {\em et~al.},
\newblock Mon. Not. Roy. Astron. Soc. {\bf 441}, 24 (2014).

\bibitem{capozziello:07}
S.~Capozziello and S.~Tsujikawa,
\newblock Phys.Rev. {\bf D77}, 107501 (2008).

\bibitem{lombriser:14c}
L.~Lombriser and J.~Peñarrubia,
\newblock Phys.Rev. {\bf D91}, 084022 (2015).

\bibitem{schmidt:09}
F.~Schmidt, W.~Hu and M.~Lima,
\newblock Phys. Rev. {\bf D81}, 063005 (2010).

\bibitem{falck:15}
B.~Falck, K.~Koyama and G.-B. Zhao,
\newblock JCAP {\bf 1507}, 049 (2015).

\bibitem{dvali:00}
G.~Dvali, G.~Gabadadze and M.~Porrati,
\newblock Phys.Lett. {\bf B485}, 208 (2000).

\end{thebibliography}

\end{document}